\documentclass[aps,prx,superscriptaddress,twocolumn]{revtex4-2}  
\usepackage{graphicx} 
\usepackage{braket}
\usepackage{bm}        
\usepackage{amssymb}   
\usepackage{amsmath}   
\usepackage{url}
\usepackage{soul,xcolor}

\usepackage{tikz,xcolor,hyperref}
\definecolor{lime}{HTML}{A6CE39}
\DeclareRobustCommand{\orcidicon}{
    \hspace{-2mm}
	\begin{tikzpicture}
	\draw[lime, fill=lime] (0,0) 
	circle [radius=0.16] 
	node[white] {{\fontfamily{qag}\selectfont \tiny ID}};
	\draw[white, fill=white] (-0.0625,0.095) 
	circle [radius=0.007];
	\end{tikzpicture}
	\hspace{-2.5mm}
}
\foreach \x in {A, ..., Z}{\expandafter\xdef\csname orcid\x\endcsname{\noexpand\href{https://orcid.org/\csname orcidauthor\x\endcsname}
			{\noexpand\orcidicon}}
}

\newcommand{\citeasnoun}[1]{Ref.~\onlinecite{#1}}

\begin{document}

\title{Inverse design of multiresonance filters via quasi-normal mode theory}

\author{Mo Chen\orcidA} \affiliation{Department of Mathematics, Massachusetts Institute of Technology, Cambridge, MA 02139, USA.}
\author{Steven G. Johnson\orcidB} \affiliation{Department of Mathematics, Massachusetts Institute of Technology, Cambridge, MA 02139, USA.}
\author{Aristeidis Karalis\orcidC} \email[Corresponding author: ]{aristos@mit.edu} \affiliation{Research Laboratory of Electronics, Massachusetts Institute of Technology, Cambridge, MA 02139, USA.}


\begin{abstract}
We present a practical methodology for inverse design of compact high-order/multiresonance filters in linear passive 2-port wave-scattering systems, targeting any desired transmission spectrum (such as standard pass/stop-band filters). Our formulation allows for both large-scale topology optimization and few-variable parametrized-geometry optimization. It is an extension of a quasi-normal mode theory and analytical filter-design criteria (on the system resonances and background response) derived in our previous work. Our new optimization-oriented formulation relies solely on a scattering solver and imposes these design criteria as equality constraints with easily calculated (via the adjoint method) derivatives, so that our algorithm is numerically tractable, robust, and well-suited for large-scale inverse design. We demonstrate its effectiveness by designing 3rd- and 4th-order elliptic and Chebyshev filters for photonic metasurfaces, multilayer films, and electrical $LC$-ladder circuits.
\end{abstract}

\maketitle

\section{Introduction}

The ability to design 2-port scattering systems with a prescribed multiresonant frequency response (``high-order filters'') had traditionally been limited to systems with separable components and of sizes that are either very subwavelength (using circuit theory~\cite{liao2020quasi,li2013synthesis,costa2014overview,sarabandi2007frequency,bayatpur2008single,nakabayashi2009band,mesa2015simplified}) or very large (via coupled-mode theory / CMT~\cite{Haus1984,fan2002analysis,suh2004temporal}). For compact wavelength-sized filters, more recent attempts at ``brute-force'' optimization of the transmission spectrum (via inverse design over a large number of degrees of freedom) have been successful in maximizing transmission over a frequency range~\cite{Chen24, Molesky2018, Hammond2022, Lu2012couplers, Frellsen2016,Callewaert2016,Shang2023}, but they have faced severe difficulties (too many local optima and convergence stiffness) when steep transitions between the passband and stopbands are required~\cite{manara1999frequency,kern2005design,bossard2006design,aage2017topology,jensen2005topology,jiang2013tailoring}, which is only achievable with structures supporting multiple sharp resonances. Fortunately, our recent work~\cite{BenzaouiaJo22} developed a systematic and computationally more tractable methodology for precise filter design, based on a reformulated quasi-normal mode theory (QNMT)~\cite{BenzaouiaJo21-QNMT}. In \citeasnoun{BenzaouiaJo22}, we derived a minimal set of analytical design criteria for $N$th-order ``standard filters'' (SFs)~\cite{dimopoulos2011analog} (reviewed in Sec.~\ref{sec:review}) and relied on an eigenmode solver to optimize microwave metasurfaces for various functionalities. However, this setup suffers the limitations that a good initial topology with $N$ resonances (QNMs) has to be intuitively preselected and these $N$ poles have to be tracked during optimization, and moreover eigensolves are computationally costly and not necessarily differentiable with respect to the structural parameters~\cite{Liang2013, Men2010}. In this article, we propose a new practical high-order-filter design algorithm (Sec.~\ref{sec:method}), which still relies on the QNMT criteria of \citeasnoun{BenzaouiaJo22}, but uses only a differentiable frequency-domain scattering solver that does not involve pole tracking and does not require a good initial structural guess, so it is suitable not only for traditional few-parameter geometry optimization but also for large-scale topology-optimization (TopOpt) methods~\cite{jensen2011topology, Maute13}. We illustrate our approach by designing 3rd- and 4th-order elliptic photonic filters with two-dimensional (2D) density-based topology optimization~\cite{jensen2011topology, Maute13}(Sec.~\ref{sec:2D-TopOpt}) as well as Chebyshev filters with simpler few-parameter optimizations (Sec.~\ref{sec:FewParam-Opt}).

The design of transmission filters with specific tolerance requirements for the pass (transmit) and stop (reflect) frequency bands is a very old problem, for which signal-processing theory provides optimal analytical solutions for the spectral response in terms of rational functions [known as ``standard filters'' (SFs), such as Butterworth, Chebyshev, inverse Chebyshev, and elliptic], with specific poles and zeros~\cite{dimopoulos2011analog}. In the quasistatic limit (wavelength $\gg$ device size), structures can be modeled as circuits of discrete elements (e.g., resistors, capacitors, and inductors or operational amplifiers in electronics) and there exist simple recipes for the appropriate circuit topologies and element values to physically implement these SFs~\cite{dimopoulos2011analog}. In the other limit of large structures (wavelength $\ll$ device size), one can use separate, cascaded, weakly coupled resonators (e.g., a chain of photonic ring resonators) and employ a coupled-mode theory (CMT) model to match an SF rational function by fine-tuning the resonators' frequencies and their nearest-neighbor couplings~\cite{little1997microring,fan1998channel,manolatou1999coupling,popovic2006multistage,xiao2007highly,liu2011synthesis,dai2011circuit}. However, in the intermediate regime of device size on the order of wavelength (ideal for compact wave filters), where structures involve strongly coupled and spatially overlapping resonances, systematic filter design has been hampered by the lack of an analytical model able to match the SF rational transfer functions. The only attempts had relied on brute-force inverse design (e.g., topology optimization) that directly targeted the transmission spectrum.

Topology optimization (TopOpt)~\cite{jensen2011topology, Maute13} (reviewed in Sec.~\ref{sec:TopOpt}) is a powerful large-scale inverse-design approach that optimizes the performance of freeform structures over thousands of degrees of freedom, and can yield  performance~\cite{Molesky2018} that would be unattainable by traditional ``intuitive'' design methods based on tuning a few carefully chosen parameters in a pre-determined geometry. In density-based TopOpt, a design domain is discretized [e.g., with a finite-element method (FEM)] and the material index at each discretization element becomes an unknown parameter~\cite{jensen2011topology, Maute13}. The objective function is usually directly related to the physical quantity of interest~\cite{Chen24}, such as the field intensity at a point (e.g., for designing the focus of a lens~\cite{Lin2018metaoptics, Christiansen20, Chung20, Li2022review}), the power exerted by a dipole source at a single frequency (for designing a resonant cavity~\cite{Liang2013,WangEl2018,Isiklar2022}), or the power transmission through the structure over a broad frequency range (for designing waveguide couplers~\cite{Lu2012couplers,Frellsen2016,Callewaert2016,Hammond2022,Shang2023,Molesky2018}). In this latter case, the optimization was commonly formulated by directly maximizing the (minimum or mean) transmission at a finite sampling of desired frequencies. However, in order to design efficient narrowband filters, where a sharp transition is required between the high- and low-transmission frequency ranges (called ``passbands'' and ``stopbands'' in filter theory), such a direct brute-force approach leads to severe numerical difficulties (Sec.~\ref{sec:naive}). The highly oscillatory nature of the desired transmission spectrum, arising from the necessary underlying sharp resonances, creates a sea of local optima that can trap the optimization in poor solutions (illustrated with an example in Fig.~\ref{fig:3rd_spectrum}). Moreover, stringent criteria on passband and stopband tolerances lead to stiff optimization problems with very slow convergence~\cite{aage2017topology,jensen2005topology}. 

To address these issues, we recently derived simple and general analytical rules to design standard filters~\cite{BenzaouiaJo22} based on our earlier introduced quasi-normal mode theory (QNMT) for the scattering matrix $S$~\cite{BenzaouiaJo21-QNMT}. In particular, we showed that the sharp (of high quality factor $Q$) 2-port-system resonances, apart from having eigenfrequencies matching the poles of the desired SF, must also have eigenfields coupling to the two ports with specific unitary ratios. In addition, an appropriate amount of background transmission must be enforced to generate the desired overall transmission spectrum away from the resonances. Essentially, the success of the method relies on the fact that, through QNMT, the rapidly oscillating part of the spectrum is designed by setting objectives on the resonances rather than the transmission itself, so the remaining slowly varying background term can be optimized more easily. A review of our QNMT~\cite{BenzaouiaJo21-QNMT} and filter-design criteria~\cite{BenzaouiaJo22} is presented in Sec.~\ref{sec:review}. Still, it is not immediately evident how to incorporate these analytical criteria into a numerically tractable formulation of an optimization problem. Previous TopOpt efforts to design cavity resonances typically maximized the power exerted by a dipole source, or equivalently the local density of states (LDOS), at a single real excitation frequency~\cite{Liang2013,WangEl2018,Isiklar2022}. However, such an approach creates only one resonance and only at a frequency with real part equal to the excitation frequency, without being able to control the QNM's decay rate (imaginary part of the pole) nor its field profile (thus its coupling to the ports). Some other works optimized for resonances by instead targeting a desired transmission spectrum, but were limited to only a single resonance~\cite{didari2024inverse} or resonances at several spectrally well-separated frequencies (non-overlapping peaks)~\cite{he2021deterministic}. Our previous work where the analytical criteria were derived~\cite{BenzaouiaJo22} utilized an eigenvalue solver to directly optimize the $N$ high-$Q$ resonances for an $N$th-order filter. However, for a random initial structure, it is not clear which $N$ of the initial (likely low-$Q$) eigenvalues are to be optimized. Similarly, it is not always easy to guess a reasonable initial structure with $N$ high-$Q$ resonances in the vicinity of the desired targets (in \citeasnoun{BenzaouiaJo22}, we used intuition from circuit theory to guess a reasonable topology and its initial geometrical parameters). Furthermore, even if $N$ high-$Q$ poles are identified, as the optimization progresses and they shift, one needs to track them smoothly (differentiably) and this is not always practical, especially when two poles cross in the complex plane. Finally, compared to scattering solvers, eigensolvers tend to be slow, since they rely on solving a sequence of many linear systems~\cite{Trefethen97}.

To overcome such difficulties, we propose a new method (described in Sec.~\ref{sec:method}) for simultaneously enforcing multiple resonances with exact desired complex poles and coupling ratios to the ports, by equivalently optimizing for the poles' conjugates, namely the $S$-matrix/coherent-perfect-absorption (CPA) zeros~\cite{krasnok2019anomalies, chong2010coherent}: the design region is excited from both ports at the conjugates of the fixed desired pole frequencies---using a \emph{complex}-frequency scattering solve---and with the conjugates of the desired coupling ratios, and then all outgoing fields are minimized to zero (see Sec.~\ref{sec:zeros}). Afterwards, to additionally obtain a background transmission with the desired value, we estimate the background matrix $C$ via a simple model requiring only calculation of the scattering through the structure at a set of real-frequency samples, and constrain $C$ at these frequencies (as described in Sec.~\ref{sec:background}). Therefore, our optimization formulation relies on only a frequency-domain scattering solver, which leads to fast calculations, fully differentiable objectives, no pole tracking, and, most importantly, no requirement of a good initial structural guess (e.g.~either a random pattern or a chosen parametrized geometry as desired). 
Although many optimization algorithms could potentially be applied to our formulation, we show that our criteria can be interpreted as a system of nonlinear equations (seeking a \emph{root} rather than a minimum, as explained in Sec.~\ref{sec:nonlinear}). For inverse design, this system is \emph{under}determined, so we propose (in Appendix~\ref{appendix:LMA}) a modified version of the Levenberg--Marquardt algorithm~\cite{fletcher2000practical} (traditionally applied to \emph{over}determined least-square problems) that works well on our examples.

Our methodology is widely applicable to practically any desired transmission spectrum and is independent of the underlying physics (it can be used for mechanical, acoustic, electromagnetic, or quantum filters). To illustrate its effectiveness, we designed narrow bandpass standard filters with large free spectral range: in Sec.~\ref{sec:2D-TopOpt}, using 2D density-based TopOpt, dielectric photonic 3rd- and 4th-order elliptic filters (the latter demonstrating the case of nontrivial background transmission), and also, in Sec.~\ref{sec:FewParam-Opt}, using few-parameter optimization, a 1D (layered) photonic 3rd-order Chebyshev filter with asymmetric ports (transmission from air to a substrate) and an electrical $LC$-ladder 4th-order Chebyshev filter. Furthermore, Appendix~\ref{appendix:Fit} describes how to fit an arbitrary spectrum to a QNMT model for use with our inverse-design algorithm.

\section{Quasi-normal mode theory and inverse-design criteria}
\label{sec:review}

\begin{figure}[ht!]
    \centering
    \includegraphics[width=1\columnwidth,keepaspectratio]{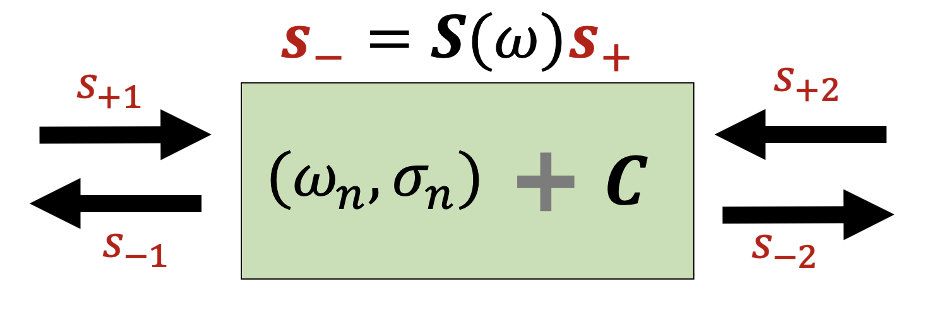}
    \caption{A schematic of a 2-port system, with input and output wave amplitudes $s_{\pm p}$ respectively, related at frequency $\omega$ by a scattering matrix $S(\omega)$, which is computed by QNMT via the high-$Q$ modes with frequencies $\omega_n$ and port-coupling ratios $\sigma_n$, plus an effective background response $C$.}
    \label{fig:QNMT}
\end{figure}

The QNMT developed in \citeasnoun{BenzaouiaJo21-QNMT} considers a multiport scattering system, whose outgoing waves are related to the incoming excitations at frequency $\omega$ by a scattering matrix $S(\omega)$, and provides a simple analytical model for calculating $S(\omega)$ with very high accuracy using only knowledge about the system resonant modes. The $N$ system high-$Q$ resonances (QNMs) have complex frequencies $\omega_n$, which appear as poles of $S$, and complex coupling coefficient to each port $p$ equal to $D_{pn}$, which can be computed as an overlap surface integral~\cite{Marcuse74} between the $n$-QNM field and the $p$-port mode at the boundary of the scatterer, as explained in detail in \citeasnoun{BenzaouiaJo21-QNMT}. Conveniently, the ratios $D_{pn}/D_{qn}$ do not depend on the normalization of the QNMs (\citeasnoun{BenzaouiaJo21-QNMT}, Appendix~D). QNMT factors these $N$ high-$Q$ resonances into a pole-expansion sum, indicated as $\bar{S}(\omega)$ below. Additionally, low-$Q$ resonances (with poles $\omega_n^C$ and couplings $\sigma_n^C$) can be factored into an effective-background matrix $C(\omega)$, which varies slowly with frequency (approximately constant in most practical design cases). For the case of a 2-port system, the ratios $\sigma_n = D_{2n}/D_{1n}$ were defined and a schematic is shown in Fig.~\ref{fig:QNMT}. The key result from \citeasnoun{BenzaouiaJo21-QNMT} is that, for a lossless system, by enforcing energy conservation (unitary $S$) and maximizing reciprocity (fine-tuning $\sigma_n$ to also best symmetrize $S$ --- exactly symmetric when $C$ constant), the $2\times2$ scattering matrix $S(\omega)$ has a simple, approximate, but very accurate expression that only depends on $\{\omega_n, \sigma_n\}$ and $C$:
\begin{equation}\label{eq:QNMT}
\begin{gathered}
S(\omega)\approx\bar{S}(\omega)\cdot C(\omega)\\ 
\bar{S}_{\{\omega_{n},\sigma_n\}} (\omega) =I+\sum_{n=1}^{N}\frac{\bar{S}^{(n)}}{i\omega-i\omega_{n}}\\
\bar{S}^{(n)}_{pq}=\sigma_{pn}\sum_{\ell=1}^{N}(M^{-1})_{n\ell}\sigma_{q\ell}^{*},\\  M_{n\ell}=\frac{1+\sigma_{\ell}\sigma_{n}^{*}}{i\omega_{\ell}-i\omega_{n}^{*}},\;\;
\left(\begin{array}{l}\sigma_{1n}=1\\\sigma_{2n}=\sigma_{n}\end{array}\right).
\end{gathered}
\end{equation}
(Here, $\bar{S}$ is defined as the indicated sum; note that a bar $\bar{\square}$ does \emph{not} denote complex conjugation of $\square$, for which we use an asterisk~$\square^*$.)
The background $C$ is itself unitary symmetric and could conceptually be computed from the remaining (presumably low-$Q$) system poles: $C(\omega)=-\bar{S}_{\{\omega_n^C,\sigma_n^C\}}(\omega)$. When no separate background is needed (all modes are considered as high-$Q$), then $C=-I$. Moderate losses can also be incorporated perturbatively~\cite{BenzaouiaJo21-QNMT} by adjusting the poles only in the denominator of the $\bar{S}$ expansion in Eq.~(\ref{eq:QNMT}).

Apart from precise forward calculation, another major benefit of this QNMT model is that it also enables the \emph{inverse} design of a lossless 2-port scattering system with practically \emph{any} desired (causal) multiresonance spectral response (commonly known as a frequency filter). Traditional design methods attempted to directly optimize for the spectral response $S(\omega)$ and faced major difficulties due to the many local optima associated with the fast oscillatory nature of a multiresonant $S(\omega)$ and due to convergence ``stiffness'' arising from stringent passband or stopband tolerances. Our QNMT method~\cite{BenzaouiaJo22} circumvents these problems, by extracting the fast oscillations into $\bar{S}(\omega)$, which is optimized indirectly by setting the QNMT parameters $\{\omega_n, \sigma_n\}$ as objectives, and thus reducing any additional needed direct optimization to the slowly varying $C(\omega)$, which is computationally more tractable.

Specifically, the first design step is to identify the (typically few) target QNMT parameters that give the desired frequency response $\tilde{S}(\omega)$ via Eq.~(\ref{eq:QNMT}), namely $N$ complex pairs $\{\tilde{\omega}_n, \tilde{\sigma}_n\}$ of high-$Q$ resonances and a slowly varying unitary symmetric $\tilde{C}=\begin{pmatrix}\tilde{r} & \tilde{t} \protect\\\tilde{t} & -\tilde{r}^*\end{pmatrix}e^{i\theta}$, where $\tilde{t}=\sqrt{1-|\tilde{r}|^2}$ is real and the overall phase $\theta$ of $\tilde{C}$ is unimportant for amplitude (namely $|S_{21}(\omega)|$) filters. Note that, under the rotating-wave approximation (RWA) where negative-frequency poles are ignored, if one adds a common phase $\varphi$ to all $\sigma_n$, the $\bar{S}$ expansion in Eq.~(\ref{eq:QNMT}) has the property $\bar{S}_{pp\{\omega_{n},\sigma_ne^{i\varphi}\}}=\bar{S}_{pp\{\omega_{n},\sigma_n\}}$, $\bar{S}_{12\{\omega_{n},\sigma_ne^{i\varphi}\}}=e^{-i\varphi}\bar{S}_{12\{\omega_{n},\sigma_n\}}$, and $\bar{S}_{21\{\omega_{n},\sigma_ne^{i\varphi}\}}=e^{i\varphi}\bar{S}_{21\{\omega_{n},\sigma_n\}}$~\cite{BenzaouiaJo21-QNMT}. Therefore, since $S=\bar{S}\cdot C$, for a target set of $\{\tilde{\omega}_n, \tilde{\sigma}_n\}$ and $\tilde{r}$, which gives the desired $\tilde{S}(\omega)$, all sets $\{\tilde{\omega}_n, \tilde{\sigma}_ne^{i\varphi}\}$ and $\tilde{r}e^{-i\varphi}$ lead to the same \emph{amplitude} and \emph{phase-delay} response for any choice of $\varphi$.

For an arbitrary desired frequency response $\tilde{S}(\omega)$, one could determine the target QNMT parameters by fitting $\tilde{S}(\omega)$ with the QNMT model of Eq.~(\ref{eq:QNMT}). Because $\tilde{S}(|\omega/\omega_n-1|\gg1)\rightarrow \tilde{C}$, one can usually infer $|\tilde{r}|$ by the desired response's asymptotic amplitude away from the resonant features, at the edges of the frequency range of interest. The procedure for this fit is detailed in Appendix~\ref{appendix:Fit}.

For particular spectral responses, however, that correspond to standard $N$th-order filters (Butterworth, Chebyshev, inverse Chebyshev, and elliptic), described by rational functions of $i\omega$ with real coefficients~\cite{dimopoulos2011analog}, \citeasnoun{BenzaouiaJo22} has shown that the target QNMT parameters conveniently take specific analytical values: (i) The $N$ high-$Q$ complex poles $\tilde{\omega}_n$ are immediately available from the denominator of the desired-filter rational function~\cite{dimopoulos2011analog}, (ii) their coupling ratios $\tilde{\sigma}_n$ must have unity magnitude and alternate signs,
\begin{equation}\label{eq:target_sigma}
    \tilde{\sigma}_n = e^{i\varphi}(-1)^{n-1},
\end{equation}
and (iii) the background response $\tilde{C}$ is constant throughout the frequency range with
\begin{equation}\label{eq:target_r}
    \tilde{r}=e^{-i\varphi}i^{N\mp1}|\tilde{r}|,
\end{equation}
with the $\mp$ signs corresponding to bandpass and bandstop filters respectively, and $|\tilde{r}|$ is 1 for an odd-order bandpass transmission SF (asymptotes to full reflection in stopband), 0 for odd-order bandstop, large (close to 1) for even-order elliptic or inverse-Chebyshev bandpass, and small (close to 0) for even-order elliptic or inverse-Chebyshev bandstop. As explained earlier, for practical design problems where the RWA applies, $\varphi$ can be any phase and is thus a design degree of freedom, or alternatively a hyperparameter that may be chosen to generate different solutions.

Once the target set of QNMT parameters is defined, one must use some optimization method to enforce them on the system. In the following, quantities $\tilde{\square}$ decorated with~$\sim$ denote the desired \emph{target} values, and quantities $\square$ without~$\sim$ denote the \emph{actual} values for the structure during the iterative optimization. That is, the optimization attempts to accomplish
\begin{equation}\label{eq:objectives}
\begin{gathered}
    \{\omega_n,\sigma_n\} \rightarrow \{\tilde{\omega}_n,\tilde{\sigma}_n\}\\
    \text{and } C_{11}^*(\omega)C_{21}(\omega) \rightarrow \tilde{r}^*\tilde{t} \, .
\end{gathered}
\end{equation}
In the last line, the $C_{11}^*C_{21}$ construction cancels out the overall unimportant phase ($\theta$) of $C$ and is computationally appropriate for both bandpass or bandstop filters (as opposed to $C_{11}/C_{21}\rightarrow \tilde{r}^*/\tilde{t}$, which would be ill-behaved for bandpass as $|\tilde{C}_{21}|=\tilde{t}\ll1$), and ${\omega}$ refers to all frequencies within the desired ``free spectral range'' (FSR) of the filter (namely the entire frequency range of interest, covering both the passband and stopband). What is then the best method to enforce this set of filter-design \emph{equality} criteria, Eqs.~(\ref{eq:objectives})?

\section{Differentiable enforcement of inverse-design criteria}
\label{sec:method}

\subsection{Enforcing multiple complex poles and~coupling~ratios}
\label{sec:zeros}

Our method for precisely designing multiple complex poles and coupling ratios, namely the first condition of Eq.~(\ref{eq:objectives}), relies on the principle of time-reversal symmetry. If a system has an $S$-matrix pole at complex $\omega_n$ that couples to ports with ratio $\sigma_n$, then it also has an ``$S$-matrix zero" at frequency $\omega_n^*$ that couples to the ports with ratio $\sigma_n^*$, and vice versa. (Note that an ``$S$-matrix zero" is not to be confused with a zero of an $S$-matrix \emph{element}, such as a transmission zero $\omega_{tz}$, where $S_{21}(\omega_{tz})=0$; instead, it refers to the complex-frequency points where the $S$-matrix itself becomes singular, i.e. $\det(S)=0$.) While the eigenfield of a QNM outside the structure is purely outgoing (with exponentially increasing tails), the field of a ``zero" is (by time-reversal) purely incoming to the structure. Therefore, in order to create a pole at $\tilde{\omega}_n$ with coupling ratio $\tilde{\sigma}_n$, we can equivalently optimize for an ``$S$-matrix zero" by exciting the two ports at $\tilde{\omega}_n^*$, where the excitation amplitudes have ratio $\tilde{\sigma}_n^*$ (i.e., $s_{+1}=1$ and $s_{+2}=\tilde{\sigma}_n^*$), and minimizing to zero the two outgoing fields ($s_{-1}$ and $s_{-2}$). By simultaneously following this procedure for all $N$ target poles $\tilde{\omega}_n$ and coupling ratios $\tilde{\sigma}_n$, we end up with a system of $2N$ complex equations for the scattering matrix elements that need to be enforced:
\begin{equation}\label{eq:zeros}
    \begin{split}
        S_{11}(\tilde{\omega}_n^*)+\tilde{\sigma}_n^*S_{12}(\tilde{\omega}_n^*)&=0 \\
        S_{21}(\tilde{\omega}_n^*)+\tilde{\sigma}_n^*S_{22}(\tilde{\omega}_n^*)&=0
    \end{split}
\end{equation}

This method uses only a frequency-domain scattering solver, at \emph{complex} excitation frequencies (which are somewhat unusual but pose no special difficulty). One key advantage of this setup, compared to using an eigenmode solver and directly optimizing $\{\omega_n, \sigma_n\}$~\cite{BenzaouiaJo22}, is that no selection of $N$ initial poles nor their tracking during optimization is needed. Another benefit is that the scattered outgoing fields, thus these conditions, are fully differentiable with respect to the structural parameters, and the derivatives can be obtained efficiently by a standard adjoint method~\cite{Molesky2018,niederberger2014sensitivity}.

\subsection{Enforcing background matrix}
\label{sec:background}

As discussed already, to acquire the desired spectral response, it is imperative for the system to also exhibit a particular background response $C(\omega)$ for real frequencies within the FSR of the filter; specifically, the last condition of Eq.~(\ref{eq:objectives}) must also be satisfied. Under our QNMT formulation of Eq.~(\ref{eq:QNMT}), we can write the background matrix as $C(\omega)\approx\bar{S}_{\{\omega_n, \sigma_n \}}^{-1}(\omega)\cdot S(\omega)$. During optimization, the same frequency-domain scattering solver we used for the poles can also compute $S(\omega)$. However, computation of $\bar{S}$ requires the knowledge of the actual $\{\omega_n, \sigma_n \}$ of the system at each optimization iteration. To find those, we could use an eigensolver; however, as we have already explained, we instead need a computationally cheap and differentiable way to compute $C$.

In the previous subsection, we presented an approach to enforce multiple complex poles and coupling ratios---that is, if this algorithm succeeds, we know that the actual resonances $\{\omega_n, \sigma_n \}$ of the system will eventually converge to their prescribed target values $\{ \tilde{\omega}_n, \tilde{\sigma}_n \}$. Consequently, $\bar{S}_{\{\omega_n, \sigma_n \}}(\omega)$ will converge to the target $\bar{S}_{\{ \tilde{\omega}_n, \tilde{\sigma}_n \}}(\omega)$, which is a known, fixed function (independent of the design parameters). Therefore, we can follow a two-step approach, where we first design the poles and coupling ratios, and, once convergence is adequate, we approximate the background by $C(\omega)\approx\bar{S}_{\{ \tilde{\omega}_n, \tilde{\sigma}_n \}}^{-1}(\omega)\cdot S(\omega)$ and use optimization to push this $C$ to its target value~$\tilde{C}$. In practice, instead of two separate steps, we enforce both criteria simultaneously in one stage, but use a small multiplier $\alpha\ll 1$ to deprioritize the constraints for the background relative to the constraints for the resonances (see Sec.~\ref{sec:nonlinear}). In this way, in the initial steps of the optimization, the former do not contribute significantly to the total objective and the focus is on the resonances. As the optimization iterates and $\{ \omega_n, \sigma_n \} \to \{ \tilde{\omega}_n, \tilde{\sigma}_n \}$, the resonances' errors are reduced and the $C$ constraints become equally significant, but then $\bar{S}_{\{ \omega_n, \sigma_n \}} \to \bar{S}_{\{ \tilde{\omega}_n, \tilde{\sigma}_n \}}$, so the $C$ approximation is indeed valid.

To constrain our approximate $C(\omega)$ to the desired value, we sample it at a number $M$ of real frequencies $\{\omega_m\}$ within the spectrum of interest and enforce the constraint of Eq.~(\ref{eq:objectives}), leading to $M$ complex equations 
\begin{equation}\label{eq:C-constraint}
    C_{11}^*(\omega_m) \cdot C_{21}(\omega_m) = \tilde{r}^*\tilde{t},
\end{equation}
where $C_{p1}(\omega_m) = \sum_{q=1,2} \bar{S}_{\{ \tilde{\omega}_n, \tilde{\sigma}_n \}pq}^{-1}(\omega_m)S_{q1}(\omega_m)$ from above.

The $M$ frequencies can be chosen to be a uniform distribution over the desired spectral range or some other set. For example, one may decide to avoid frequencies close to the target poles, where the $C\approx\bar{S}^{-1}\cdot S$ approximation may entail larger errors.

In principle, successfully enforcing Eqs.~(\ref{eq:C-constraint}) at a finite set of frequencies might not always suffice to obtain the desired $C(\omega)$ at \emph{all} frequencies of interest. The reason is that very-high-$Q$ modes may still manage to ``fit between'' two sample frequencies $\omega_m$ and create in the response a resonant spike between those frequencies. Obviously, increasing $M$ could reduce these occurrences, but that would make each iteration very costly computationally and it is still not a guarantee (modes with arbitrarily high $Q$ can still appear). In practice, however, we found that, for a reasonable desired FSR, often a small number $M<100$ of sample frequencies was sufficient to produce good results. Also, when an optimization result included unwanted modes in the background response close to the edges of the desired FSR, we saw that a subsequent (second-step) optimization with the $M$ frequencies just inside these unwanted modes could sometimes manage to push them away from the filter bandwidth and increase the FSR. 

Another optional tool at the designer's disposal to reduce the number of undesired high-$Q$ modes is to additionally limit the total amount of material in the structure, since the density of modes increases with the material volume. If $\rho_\text{p}(\mathbf{r})$ denotes the (dimensionless) TopOpt ``density'' parameter (ranging between 0 and 1) at each location $\mathbf{r}$ in the design region $V$, the total material volume is $V_\rho = \int_V \rho_\text{p}(\mathbf{r})dV$. Then, this additional limit can be written as $V_\rho \leq V_{\rho,\text{max}}$ or, by using a slack variable $z$ to convert the inequality constraint to an equality constraint plus bounds for the variable,
\begin{equation}\label{eq:n-constraint}
    V_\rho = z; \;\; 0 \leq z \leq V_{\rho,\text{max}}.
\end{equation}
If used, this condition should also be scaled by another multiplier $\gamma>1$, which should be large enough to ensure its enforcement (say $\gamma=10$). Alternatively, one could minimize $V_\rho$ under the equality constraints of Eqs.~(\ref{eq:zeros}) and (\ref{eq:C-constraint}), but we did not test that option in this work.

Similarly to Eq.~(\ref{eq:zeros}), the background-enforcing conditions Eqs.~(\ref{eq:C-constraint}) and (\ref{eq:n-constraint}) are also fully differentiable with respect to the structural parameters, as is necessary for TopOpt methods.

\subsection{Design criteria as nonlinear least-squares}
\label{sec:nonlinear}

Our design method has been reduced to the system of equations, Eqs.~(\ref{eq:zeros}), (\ref{eq:C-constraint}), and optionally (\ref{eq:n-constraint}). Since each complex equation is equivalent to two real equations, these amount to $4N+2M(+1)$ real equations. That is, our optimization problem corresponds more closely to a nonlinear root-finding problem, in contrast to typical optimization approaches that involve minimization of some quantity.

Among the available algorithms, we experienced the best performance with the Levenberg--Marquardt algorithm (LMA), which aims to minimize the sum of squares of all equations through a dynamic interpolation between the Gauss--Newton and steepest-descent methods~\cite{fletcher2000practical}. LMA basically converts the nonlinear root-finding problem to a least-squares one, where minimization is indeed the goal but the minimum is hoped to be exactly zero.
In forming this sum of squares, we deprioritize the background constraints by multiplying the corresponding $2M$ equations by a multiplier $\alpha\ll 1$ and prioritize the optional $(+1)$ total-material constraint with another multiplier $\gamma>1$, as mentioned earlier. In other words, we try to minimize $\sum_n \|\text{Eq.}\,\eqref{eq:zeros}\|^2 + \sum_m \|\alpha\cdot\text{Eq.}\,\eqref{eq:C-constraint}\|^2 + \|\gamma\cdot\text{Eq.}\,\eqref{eq:n-constraint}\|^2$.

Least-squares problems are usually overdetermined, namely the number of equations is much larger than the number of unknowns. However, for our design problem, if a large-scale (many-parameter) optimization is used, then a vastly underdetermined problem arises. In these cases, the traditional formulation of LMA can lead to very large and rank-deficient matrices. Therefore, we have implemented a modification to the LMA algorithm that resolves this issue and works well for inverse design. The details are presented in Appendix~\ref{appendix:LMA}.

\subsection{Summary of optimization hyperparameters}

While we find that our proposed design methodology often gives rapid convergence to a high-performance solution, far more easily and reliably than direct optimization of the transmission spectrum, the need for intelligent guidance is not entirely eliminated. As in any non-convex optimization problem, some experimentation is still required to find good values for the ``hyperparameters'', which are the variables that affect the problem and algorithm \emph{formulation} (as opposed to the problem structural ``parameters'' sought iteratively during the optimization). In this section, we summarize our method's hyperparameters that need to be chosen by the designer, and motivate choices that are more likely to lead to good results.

We have already mentioned the phase $\varphi$, the set of $M$ frequencies where the background transmission is constrained, the optional total-material limit $V_{\rho,\text{max}}$, and the multipliers $\alpha$ and $\gamma$. There are also some hyperparameters involved in the LMA, discussed in Appendix~\ref{appendix:LMA}.

Additionally, the size of the design region is obviously a very important hyperparameter (or sometimes even an explicit optimization parameter). Larger design regions may increase the likelihood of finding a solution, but, as mentioned earlier, more high-index material leads to more undesired resonances that can pollute the desired background transmission. Therefore, we recommend starting with a small design region and increasing it as needed. A good general principle is that the size should be on the order of $N\lambda$ for an $N$th-order filter, whose FSR is centered around wavelength $\lambda$. Given a design region, the maximum corresponding number of degrees of freedom is then determined by the computational mesh resolution (which is set mainly by the desired accuracy).  

Moreover, it should be made clear that this TopOpt procedure does not require a ``good'' initial structure. Nevertheless, it does not hurt to make the optimizer's job easier: it is wise to choose an initial structure that is closer to satisfying the background transmission $\tilde{t}$ criterion. For bandpass or bandstop filters, we would therefore choose an initial structure that gives low or high transmission, respectively, over the FSR of interest. For instance, for a bandstop filter between two same ports, one might choose an initial structure that homogeneously connects the two ports, allowing full transmission. For a bandpass filter, one might start with a highly reflective mirror, such as a reflective material or a periodic bandgap structure~\cite{Brillouin1956,joannopoulos2011photonic}.

\section{Topology optimization for 2D photonic metasurface filters}
\label{sec:2D-TopOpt}

\subsection{Physical model of metasurface scattering}

We will demonstrate our proposed filter-design method in the context of two-dimensional ($xy$) photonic metasurfaces, planar purely dielectric structures of finite axial thickness $d$ (in $x$) and infinite extent laterally (in $y$) but $y$-periodic with periodicity $a$. A plane wave (propagating along $x$) is normally incident onto the metasurface and we wish the transmission spectrum to exhibit a desired filter response. 

\begin{figure}[ht!]
    \centering
    \includegraphics[width=1\linewidth]{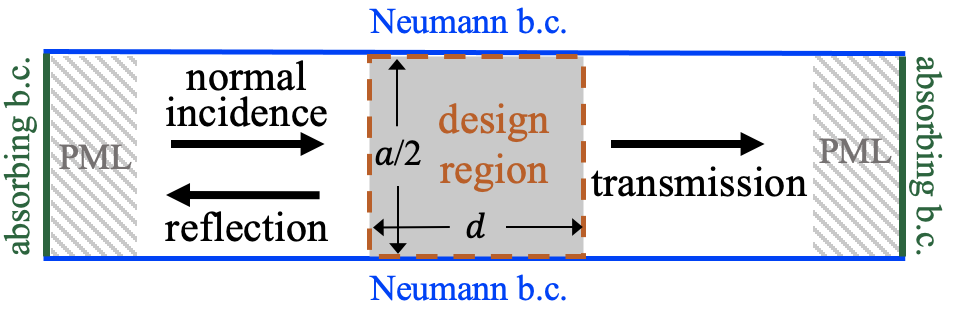}
    \caption{TopOpt setup: The design region of width $d$ and height $a/2$ is bounded vertically by Neumann boundary conditions, thus forming a photonic metasurface (infinite vertically with period $a$). The two input/output port modes are normally incident plane waves, while both PML and absorbing boundary conditions are imposed at the left/right boundaries of the computation cell to emulate free space.}
    \label{fig:TO_setup}
\end{figure}

Our computational model for this 2D setup is shown in Fig.~\ref{fig:TO_setup}. To form the metasurface, we imposed Neumann boundary condition on the bottom ($y=0$) and top ($y=a/2$) boundaries of the simulation domain, which is equivalent to mirror symmetry at both boundaries and thus a periodic boundary condition with period $a$, namely periodic unit cell with twice the height of the domain and symmetric in $y$. Note that, for this configuration, diffraction starts at free-space wavelengths smaller than the period $a$ and those diffraction orders would become loss channels; therefore, one should keep the period $a$ smaller than the smallest wavelength within the desired free spectral range (FSR) of the filter. On the left and right $x$-ends of the simulation domain, \emph{both} perfectly matched layers (PMLs)~\cite{BEREGNER_1994,Sacks1995PML, OSKOOI2011PML} and normal-incidence absorbing boundary conditions~\cite{Belhora1995} are imposed to emulate scattering into free space (no reflections from these boundaries) and reduce spurious modes from numerical discretization (see \citeasnoun{BenzaouiaJo21-QNMT}, Appendix~F for details). In all examples presented in this work, we consider the wave polarization with the $H$-field out of the 2D plane (along $z$), and thus we solve the following Maxwell's equation in frequency domain 
\begin{equation}
\label{eq:Maxwell}
    [-\nabla \cdot\frac1{\varepsilon(x,y)}\nabla-k_o^2\mu]H_z = J
\end{equation}
where $\varepsilon(x,y)$ describes the 2D relative dielectric permittivity distribution, the relative magnetic permeability here is always assumed to be $\mu=1$, $k_o=\omega\sqrt{\varepsilon_o\mu_o}$ is the wave vector in free space, and $J$ is a source term that describes the incident plane wave.

Both constraints of Eqs.~(\ref{eq:zeros}) and (\ref{eq:C-constraint}) require computing the amplitudes of the outgoing waves $s_{-p}$ ($p=1,2$), by evaluating the overlap surface integral~\cite{Marcuse74} between the field and the $p$-port mode at the closest cross section of the port to the scattering filter (see discussion in \citeasnoun{BenzaouiaJo21-QNMT} for this proper choice of cross sections). For our metasurface structure with normally propagating plane waves, they can be calculated as $s_{-p}=\int_{0}^{a/2}H_z^\text{scat}(x_p,y) dy$, where $x_1,x_2$ are the $x$-limits of the design region and $H_z^\text{scat}(x_p,y)$ is the scattered magnetic field, namely the total field minus the incident field. Subsequently, in taking the derivative of $s_{-p}$ (hence our objectives) with respect to the unknown design parameters, the adjoint method transforms the derivative of $s_{-p}$ with respect to $H_z$ into a line-source term in the adjoint equations~\cite{niederberger2014sensitivity}.

\subsection{Density-based topology optimization}
\label{sec:TopOpt}

\begin{figure*}[ht!]
    \centering
    \includegraphics[width=0.75\linewidth]{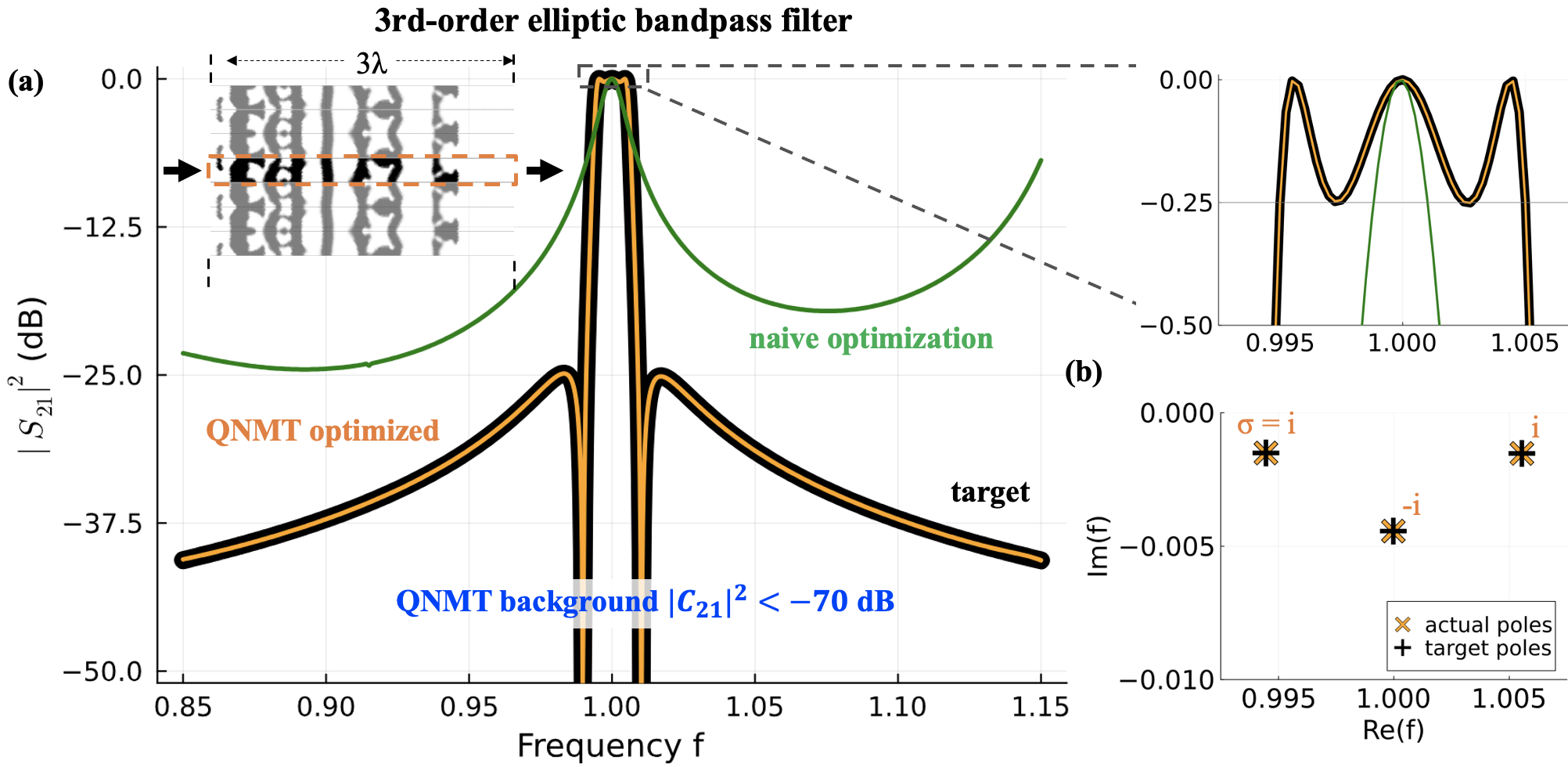}
    \caption{Designed photonic 3rd-order elliptic bandpass filter: (a) The structure is shown on the top left, with the design region boxed inside the dashed orange line. Its transmission spectrum $|S_{21}(f)|^2$ is plotted as the orange curve, with an enlargement of the passband on the top right. Its background transmission $|C_{21}(f)|^2$ is below $-70$dB (and hence omitted from the plot). Therefore, the response of our design matches perfectly the target standard filter (black curve). The traditional naive-optimization method performs very poorly in comparison (green curve). (b) Target and designed poles and their coupling ratios $\{\omega_n, \sigma_n\}$.}
    \label{fig:3rd_spectrum}
\end{figure*}

To numerically design such 2D photonic metasurface filters with our proposed methodology of enforcing the criteria Eqs.~(\ref{eq:zeros})-(\ref{eq:n-constraint}), we used an inverse-design technique of density-based TopOpt~\cite{jensen2011topology}, where every ``pixel'' in the design region is a degree of freedom. In particular, we solved Eq.~(\ref{eq:Maxwell}) with a finite-element method (FEM)~\cite{Badia2020}, where $\varepsilon(x,y)$ of each element in the design region $\Omega$ is parametrized to vary between two values $n_\text{low}^2<n_\text{high}^2$ [Eq.~(\ref{eq:nmap})] via the variables $\rho_\text{p}(x,y)$, which vary between 0 and 1 as tanh-projected~\cite{Wang2011} versions of the unbounded variables $\rho_\text{f}(x,y)$ [Eq.~(\ref{eq:proj})], which are in turn Helmholtz-filtered (through a damped diffusion PDE with radius $r_\text{f}$~\cite{lazarov2011filters}) versions of the raw design variables $\rho(x,y)$ [Eq.~(\ref{eq:helm-filter})]:
\begin{equation}\label{eq:helm-filter}
-r_\text{f}^2\nabla^2\rho_\text{f}+\rho_\text{f}=\rho ,\quad\text{with} \;
 \vec{n} \cdot \nabla \rho_\text{f} =0 \;\text{on} \; {\partial\Omega}
\end{equation}
\begin{equation}\label{eq:proj}
\rho_\text{p} =  0.5+\frac{\tanh\left[\beta_\text{p}(\rho_\text{f}-0.5)\right]}{2\tanh(\beta_\text{p}/2)}
\end{equation}
\begin{equation}\label{eq:nmap}
    \varepsilon=[n_\text{low}+\rho_\text{p}(n_\text{high}-n_\text{low})]^2
\end{equation}
The filtering (with radius $r_\text{f}=0.02\lambda$) and projection (with strength $\beta_\text{p}=8$) are used here to design more realistic structures, respectively without extremely fine features and mostly binarized between the two materials with refractive indices $n_{\text{low}}<n_{\text{high}}$, while maintaining consistency with other photonic TopOpt works~\cite{jensen2011topology,Hammond2022,YaoVe23, Chen24}. However, since this work is focused on the new filter-design method, for simplicity, we did not attempt to enforce strict length-scale constraints~\cite{sigmund2007morphology} nor to fully binarize the structure (by $\beta_\text{p}\rightarrow\infty$)~\cite{Wang2011}.

As mentioned earlier, our filter-design objectives are differentiable functions of the field $H_z$. Moreover, the derivative of the Maxwell operator Eq.~(\ref{eq:Maxwell}) with respect to $\varepsilon(x,y)$ and thus the filtered and projected design variables $\rho_\text{p}(x,y)$ can be computed. Therefore, the gradient of our objectives with $\rho_\text{p}(x,y)$ can be evaluated with the adjoint sensitivity analysis~\cite{niederberger2014sensitivity,jensen2011topology}. Finally, the Jacobian with respect to the raw design variables $\rho(x,y)$, before filtering and projection, can be obtained via the chain rule of differentiation. Consequently, the optimization problem can be solved with gradient-based algorithms. This is absolutely necessary for large-scale optimization problems, such as the one here, since the number of unknown design parameters would be too large either to compute derivatives directly with finite differences or to use derivative-free algorithms~\cite{sigmund2011usefulness}.

We generated the FEM mesh with the free/open-source software Gmsh~\cite{geuzaine2009gmsh}. We numerically solved Maxwell's equation and computed its adjoint using the free/open-source FEM package Gridap.jl~\cite{Badia2020} in the Julia language~\cite{bezanson2017julia}. Automatic differentiation was performed via the packages Zygote.jl~\cite{Zygote2018} and ChainRulesCore.jl~\cite{chainrulescore}. The Levenberg–-Marquardt algorithm (to solve the nonlinear least-squares problem) is implemented in Julia by the free open-source package LeastSquaresOptim.jl~\cite{nonlinearlstsq}, and we added the modification detailed in Appendix~\ref{appendix:LMA}. The computation was performed on the MIT Supercloud cluster~\cite{reuther2018interactive}, where the simulations at the $N+M$ frequencies $\omega_n^*$ and $\omega_m$ were performed in parallel.

\subsection{2D photonic metasurface filter examples}

The specifications of all transmission filters designed below are: bandpass type with $1\%$ pass bandwidth around $\lambda=1$, $0.25$dB transmission ripple allowed in the passband, and at least $25$dB transmission suppression in the stopband. The two dielectric materials considered for the structure are $n_\text{low}=1$ (air) and $n_\text{high}=3.4$ (silicon). The incidence and transmission air regions are $0.5\lambda$ thick, which is also the thickness of the PMLs, with absorption large enough to reduce reflections to $10^{-12}$ at the left and right boundaries. To make the optimization of this bandpass filter easier, the initial structure was always chosen to be a quarter-wave Bragg stack at $\lambda=1$, namely alternating layers of $0.25$-thick air and $0.25/n_\text{high}$-thick silicon, so that the background transmission is close to its desired \emph{small} value. To obtain reasonable computational accuracy, we chose a mesh density of about 6000 triangular mesh elements per $\lambda^2$ (so the average side of the triangular elements is roughly $0.02\lambda$, or $0.07\lambda/n_\text{high}$ in the material).


As a first example, we designed a third-order ($N=3$) elliptic filter. For an odd-order filter, $\tilde{t}=0$, so the $C$ constraint in Eqs.~(\ref{eq:objectives}) and (\ref{eq:C-constraint}) becomes $C_{21}(\omega_m)\rightarrow 0$. The design region is chosen as $3\lambda$ long and $0.25\lambda$ high ($a=0.5\lambda$). The background transmission was constrained at $M=21$ frequency points, equally spaced within $[0.85, 1.15]/\lambda$. After trying multiple values for $\alpha$ and $\varphi$, we obtained the best optimization result for $\alpha=0.02$ and $\varphi=\pi/2$, namely the coupling ratios are $\sigma_n=\{i, -i, i\}$ from Eq.~(\ref{eq:target_sigma}). The relative errors of the actual poles and coupling ratios, compared with the targets, are within $10^{-5}$. Moreover, $C_{21}$ is less than $-70$dB, effectively zero, for a free-spectral range of $30\%$. Therefore, as shown in Fig.~\ref{fig:3rd_spectrum}, we had an exceptionally good match between the actual transmission and the target spectrum.

\begin{figure}[ht!]
    \centering
    \includegraphics[width=\linewidth]{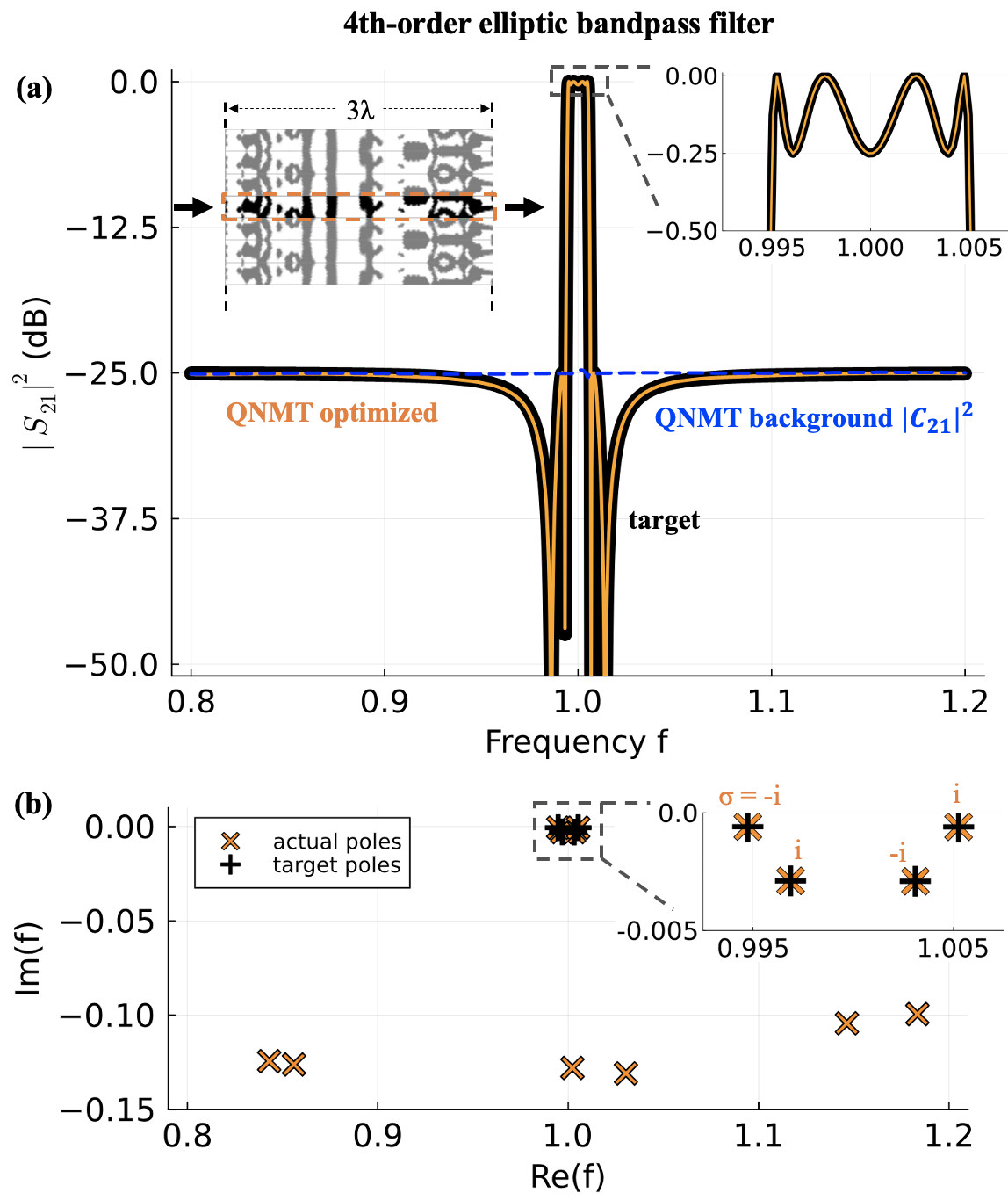}
    \caption{Designed photonic 4th-order elliptic bandpass filter: (a) The structure is shown on the top left. Its transmission spectrum $|S_{21}(f)|^2$ is plotted in orange, with an enlargement of the passband on the inset. Its background transmission $|C_{21}(f)|^2$ is steady around the specified $-25$dB. Again, our design exhibits perfect match to the target standard filter (black curve). (b) Poles of the designed structure: the high-Q poles and their coupling ratios $\{\omega_n, \sigma_n\}$ shown in the enlarged inset match their targets, while the low-Q modes collectively account for the desired background transmission.}
    \label{fig:4th_spectrum}
\end{figure}

As a second example, we designed a fourth-order ($N=4$) elliptic filter, in which (even-$N$) case the $-25$dB stopband limit sets the background transmission target to the non-trivial value $\tilde{t}=10^{-25/20}=0.0562$. We chose $\varphi=-\pi/2$, so the coupling ratios are $\sigma_n=\{-i, i, -i, i\}$ from Eq.~(\ref{eq:target_sigma}) and the target background reflection coefficient is $\tilde{r}=0.9984+\text{i}0$ from Eq.~(\ref{eq:target_r}). As before, the design region is made $3\lambda$ long and $0.25\lambda$ high ($a=0.5\lambda$). For the background constraint, we used $M=92$ frequencies equally spaced within $[0.8, 1.2]/\lambda$, with $\alpha=1/150/\sqrt{M}$. After optimization, again the actual poles and coupling ratios have reached their targets to within $10^{-5}$, while $|C_{21}|^2$ is almost exactly at $-25$dB for a $40\%$ FSR. As shown in Fig.~\ref{fig:4th_spectrum}, again we have an amazing match of the actual transmission to the target spectrum.

\subsection{Comparison to naive direct design}
\label{sec:naive}

In order to confirm the superiority of our proposed optimization method, we also attempt the conventional method of directly optimizing (for the same design region and mesh) the transmission spectrum. In particular, for the third-order elliptic filter of Fig.~\ref{fig:3rd_spectrum}, we set the optimization objective as $\min_\rho \int (|S_{21}^\text{target}|^2 - |S_{21}^\text{actual}|^2)^2d\omega $,  minimizing the least-squares difference between the target SF transmission spectrum and the actual transmission spectrum. The integration was numerically evaluated with Gauss quadrature at 95 points within the frequency range $[0.9, 1.1]/\lambda$. (The increase in points and decrease in range were in recognition of the difficulty of this approach.) The optimization was performed with the CCSA algorithm~\cite{Svanberg2002} from NLopt~\cite{nlopt}.

The result is shown with a green line in Fig.~\ref{fig:3rd_spectrum}. It can be seen that, although the curve has roughly the shape of a bandpass filter, the transmission in the stopband is not sufficiently low, while at the same time the passband is not sufficiently wide, with only one peak; more importantly, the optimization fails to achieve a sharp transition between passband and stopband, rendering the result useless.


\section{Few-parameter optimization}
\label{sec:FewParam-Opt}

Although TopOpt is a very powerful tool that can provide good designs for difficult problems, it leads to topologies that are often difficult to fabricate with standard processes and it can also be complicated to set up. In some cases, whether for easier manufacturability or for computational simplicity, one may attempt a design using a parametrized geometry (as was done in \citeasnoun{BenzaouiaJo22}). To demonstrate the generality of our proposed filter-design method, we also implement it for such simpler few-parameter optimization problems. For these small-scale problems, adjoint calculation of the gradient obviously helps but is not absolutely required; instead, finite-difference derivatives can also be used.

\subsection{1D-layered photonic filter}
\label{1d-example}

We consider a layered dielectric structure, and, in contrast to density-based optimization where each mesh element had fixed size and the design parameters were the relative permittivities, here we fix the permittivities of the layers and use the layer thicknesses as unknown design parameters. (This is sometimes referred to as ``shape optimization''~\cite{Maute13}.) Moreover, when a thickness converges (close) to a zero value, that material layer can be removed, and the optimization can be repeated with fewer parameters.

Such 1D problems can easily be solved with a transfer-matrix formulation~\cite{Yeh2005}, whose details are summarized in Appendix~\ref{appendix:TM} for the analytical calculation of the criteria Eqs.~(\ref{eq:zeros}), (\ref{eq:C-constraint}) and their gradients.

Since the design region is not of fixed size in this case, to contain the number of modes arising during optimization, it may be a good idea to limit not only each layer-thickness design variable $d_k\leq d_{k,\text{max}}$, but also the structure's total thickness (sum of all $K$ layers), by using a slack variable $y$:
\begin{equation}\label{eq:d-constraint}
    \sum_{k=1}^K d_k = y; \;\; 0 \leq y \leq d_{\text{max}}
\end{equation}
If desired, the high-index region can also be limited, similarly to Eq.~(\ref{eq:n-constraint}), by
\begin{equation}\label{eq:dn-constraint}
    \sum_{k=\text{high}} d_k = z; \;\; 0 \leq z \leq d_{\rho,\text{max}}(<d_{\text{max}})
\end{equation}

\begin{figure}
    \centering
    \includegraphics[width=1\linewidth]{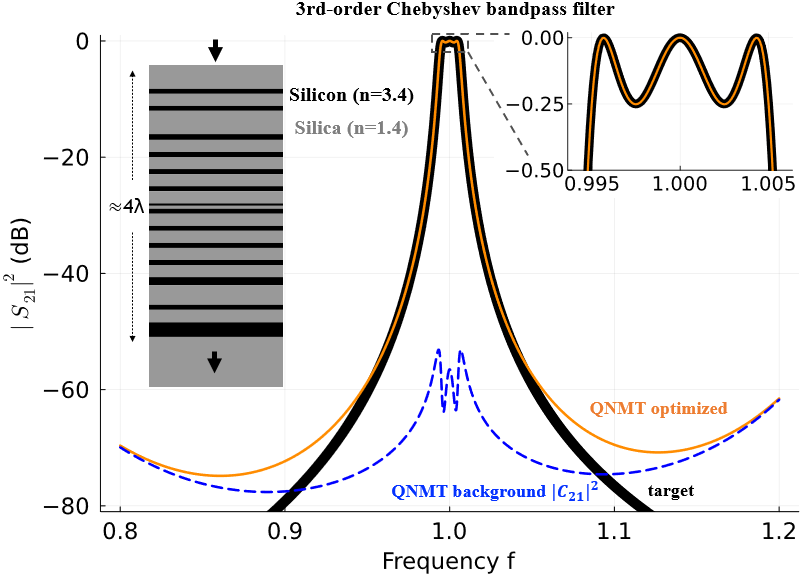}
    \caption{Designed 1D-layered photonic 3rd-order Chebyshev bandpass filter: The dielectric-layer stack is shown on the left, where black is Si and grey is SiO2, so light is normally incident from the top air and transmits into the bottom silica substrate; the $28$ layer thicknesses (from top to bottom, alternating SiO2--Si) are $d_k/\lambda=(0.3528,$ $0.07358,$ $0.1787,$ $0.07361,$ $0.3449,$ $0.08524,$ $0.1795,$ $0.07385,$ $0.1793,$ $0.07383,$ $0.1794,$ $0.07391,$ $0.1804,$ $0.03658,$ $0.04277,$ $0.07453,$ $0.1794,$ $0.07382,$ $0.1792,$ $0.07380,$ $0.1793,$ $0.07385,$ $0.1797,$ $0.1212,$ $0.2876,$ $0.07501,$ $0.1854,$ $0.2154)$. The transmission spectrum $|S_{21}(f)|^2$ (orange) matches the target filter (black) for a wide frequency range around the passband (enlarged in the inset). The background transmission $|C_{21}(f)|^2$ is below $-58$dB for the entire range.}
    \label{fig:1d_shape}
\end{figure}

As an example, we considered such a stack of dielectric films, alternating between silicon (Si with $n_\text{high}=3.4$) and silicon dioxide (SiO2 with $n_\text{low}=1.4$), and sitting on a SiO2 substrate. We designed a third-order ($N=3$) Chebyshev filter (again bandpass with $1\%$ bandwidth and $0.25$dB passband ripple) for plane-wave transmission from the air region above the stack into the substrate (assumed to be semi-infinite). We chose $\varphi=0$, namely the coupling ratios are $\sigma_n=\{1, -1, 1\}$ from Eq.~(\ref{eq:target_sigma}). [Note that this will force the tails of the QNMs to be (anti-)symmetric, although the structure is asymmetric.] We started with a 29-layer quarter-wave Bragg stack, namely $d_k=0.25\lambda/n_k$ for $k=1...29$, where $n_k$ is the refractive index of each layer, so $d_{k=\text{odd}}=0.0735$ for Si and $d_{k=\text{even}}=0.179$ for SiO2. We bounded the individual layer thicknesses to $0.75/n_k$ and the total Si thickness of Eq.~(\ref{eq:dn-constraint}) to $d_{\rho,\text{max}}=1.5N\lambda/n_\text{high}$ (with multiplier $\gamma=10$), while the total thickness of Eq.~(\ref{eq:d-constraint}) was left unconstrained. The initial Bragg stack already gives very low transmission, matching our desired background, and it turned out that we did not need to impose any background constraints ($M=0$).

The optimization led to a structure where the top Si layer was less than $0.01/\lambda$ thick. Therefore, we removed this layer and repeated the optimization with the remaining 28 thicknesses. The resulting transmission spectrum is shown in Fig.~\ref{fig:1d_shape}. Once again, an excellent match between the target and actual poles and passband transmission is observed. $C_{21}$ is less than $-53$dB, sufficiently close to zero, for a free-spectral range of over $40\%$.

As a test, we also applied our density-based TopOpt method for this problem. By reducing the height of the 2D computational cell in Fig.~\ref{fig:TO_setup} to a single pixel, the problem becomes effectively 1D. We confirmed that a similarly performing multilayered solution for this 3rd-order Chebyshev filter was indeed found.

\subsection{$LC$-ladder electronic filter}
\label{LC-example}

\begin{figure}
    \centering
    \includegraphics[width=1\linewidth]{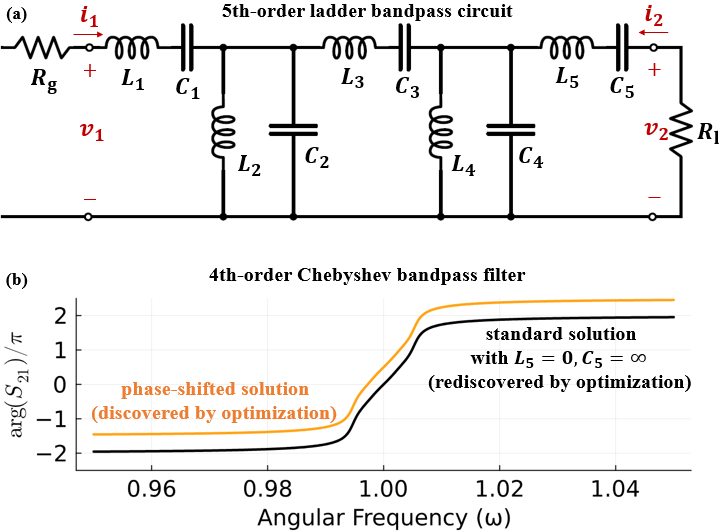}
    \caption{(a) 5th-order $LC$-ladder electrical circuit between a generator with impedance $R_\text{g}$ and a load $R_\text{l}$, with a series branch first. (b) Transmission phase for two designed 4th-order Chebyshev bandpass filters: the (textbook) standard one with $\tilde{\sigma}_1 = -i$ [using $L_k=(137.8,7.878\text{m},205.6,11.75\text{m},0)$, $C_k=(7.256\text{m},$ $126.9,$ $4.864\text{m},$ $85.10,$ $\infty)$; $\text{m}=10^{-3}$] is easily found by our optimization method, but another non-standard optimization solution with $\tilde{\sigma}_1 = 1$ [using $L_k=(137.3,$ $7.808\text{m},$ $202.0,$ $\infty,$ $222.8)$, $C_k=(7.282\text{m},$ $128.1,$ $4.991\text{m},$ $0.624,$ $4.521\text{m})$] exhibits a constant-$\pi/2$ phase-shifted response.}
    \label{fig:LC-ladder}
\end{figure}

Perhaps the simplest and most famous bandpass filter is an electronic-circuit ``ladder'' of alternating series and parallel combinations of discrete inductors $L$ and capacitors $C$, filtering the signal from a generator with impedance $R_\text{g}$ towards a load with impedance $R_\text{l}$, as shown in Fig.~\ref{fig:LC-ladder} for a 5th-order ladder starting with a series branch. To use this topology for the ``minimum-phase'' (without zeros) Butterworth or Chebyshev SFs, analytical solutions have been known for decades~\cite{dimopoulos2011analog}, but, to our knowledge, it had never been pointed out before that the overall QNMs of the textbook solutions for this simple bandpass-SF circuit have $\tilde{\sigma}_n = (-i)^{N+1}(-1)^{n-1}$ [while the topology with a parallel branch first has $\tilde{\sigma}_n = (-i)^{N-1}(-1)^{n-1}$]. Here, we will use our method to confirm these standard solutions, but also to show that this circuit supports additional SF solutions with different phase responses. (These cascaded circuits are most easily solved using the $\mathcal{ABCD}$ matrix~\cite{jarry2007rf}, as reviewed in Appendix~\ref{appendix:TM}.)

Using the 5th-order circuit of Fig.~\ref{fig:LC-ladder}, we optimized the $L$ and $C$ values for our poles criteria Eqs.~(\ref{eq:zeros}) to obtain Chebyshev bandpass filters, again with 1\% bandwidth around $\omega=1$ and minimum $-0.25$dB transmission in the passband. The criteria Eq.~(\ref{eq:C-constraint}) to set the background $C(\omega)\rightarrow 0$ were not needed. Our starting element values were chosen as $100$ for each series $L$ and parallel $C$, and $0.01$ for each series $C$ and parallel $L$, with the motivation that initially each series and parallel branch is resonant at $1/\sqrt{L_k C_k}=\omega=1$ with $Q=\omega L^\text{s}_k/R_\text{l}$ or $\omega C^\text{p}_k/R_\text{l}=100=1/$bandwidth. For a 5th-order filter, with the standard $R_\text{l}=R_\text{g}$ and $\tilde{\sigma}_1 = -1$, the result was, unsurprisingly, the textbook analytical solution. Then, we optimized for a 4th-order filter, with $R_\text{l}/R_\text{g}=2A-1+2\sqrt{A(A-1)}=1.6196$ where $A=10^{0.25/10}$~\cite{dimopoulos2011analog} and $\tilde{\sigma}_1 = -i$, and again we obtained the standard 4th-order solution with $L_5=0$ and $C_5\rightarrow\infty$ (see Fig.~\ref{fig:LC-ladder} caption): the optimization ``discovered'' that it could obtain a 4th-order solution by effectively removing part of the circuit. However, our approach also allows us to seek more non-standard solutions. In particular, we targeted the same 4th-order Chebyshev filter, but with the non-standard phase $\tilde{\sigma}_1 = 1$. We find a new solution (values in the Fig.~\ref{fig:LC-ladder} caption indicate $L_4$ is removed) that matches the desired transmission amplitude spectrum with an error of at most $0.5\%$ in $|S_{21}|^2$, but with an additional constant $\pi/2$ phase shift over the entire frequency range. Such solutions could be very useful, allowing one to combine filtering and phase-shifting functionalities.

\section{Conclusions}\label{sec:conclusion}

We have presented a new practical method for designing precise high-order spectral filters using either large-scale inverse design (TopOpt) or simple few-parameter optimization. It is based on our previously derived analytical criteria on the system resonances $\omega_n$, their coupling ratios $\sigma_n$ to ports, and the background response $C$~\cite{BenzaouiaJo22}, but, in contrast to that previous work, our present algorithm imposes these criteria using only a differentiable frequency-domain scattering solver (instead of an eigensolver) and without requiring any good initial structural guesses.

We demonstrated our approach by designing electromagnetic filters, but our method should be directly applicable to \emph{any} other linear low-loss physical system with a desired frequency-dependent response, such as for mechanical, acoustic, or quantum-scattering filters. 
We used plane-wave ports in our examples, but other common ports (e.g., waveguides) are also possible, provided the port-mode profiles vary slowly with frequency. 
Although we focused on the useful standard filters, showing implementations of 3rd- and 4th-order elliptic and Chebyshev filters, our framework can be easily extended to target \emph{any} desired transmission spectrum, by fitting it to QNMT (as described in Appendix~\ref{appendix:Fit}) to extract the relevant target QNMT parameters \{$\tilde{\omega}_n, \tilde{\sigma}_n$\} and $\tilde{C}$. 

We formulated our optimization as a nonlinear root-finding/least-squares problem and solved it with a modified version of the Levenberg-Marquardt algorithm. However, it is also possible to use a typical optimization formulation that comprises one objective function and several equality constraints (e.g.,~via sequential quadratic programming~\cite{kraft1988software,kraft1994algorithm}) and/or inequality constraints (e.g.,~via the CCSA algorithm~\cite{Svanberg2002}). For example, since the poles criteria \{$\omega_n, \sigma_n$\} must be satisfied exactly, while the background criterion $C(\omega)$ allows for some relaxation, one could first optimize only for the poles and then (as a second step) minimize the error of the background, having the poles as equality constraints; optionally, an inequality constraint for the total material or for manufacturability could be added.

\begin{acknowledgments}

The authors acknowledge the MIT SuperCloud and Lincoln Laboratory Supercomputing Center for providing high-performance-computing resources that have contributed to the research results reported within this article~\cite{reuther2018interactive}.

This work was supported in part by the Simons Foundation collaboration on Extreme Wave Phenomena, and by the U.S. Army Research Office (ARO) through the Institute for Soldier Nanotechnologies (ISN) under award no.~W911NF-23-2-0121.

\end{acknowledgments}

\bibliographystyle{unsrt}
\bibliography{Ref}

\appendix

\section{QNMT fit to a transmission spectrum}\label{appendix:Fit}

Suppose one wishes to design a 2-port structure that exhibits an arbitrary transmission-amplitude spectrum $|\tilde{S}_{21}(\omega)|^2$, e.g.~given simply as an array of frequencies and corresponding transmissions. In order to use our inverse-design method, one needs to find the target QNMT parameters $\{\tilde{\omega}_n,\tilde{\sigma}_n\}$ and $\tilde{C}$ that would generate this spectrum.

The first step is to fit this spectrum to a rational function through a rational approximation algorithm, such as Remez~\cite{trefethen2019approximation} or AAA~\cite{AAA}. This yields an expansion of the form
\begin{equation}\label{eq:targetS21}
    |\tilde{S}_{21}(\omega)|^2 = \tilde{A} + \sum_{n=1}^{N}\frac{\tilde{R}_n}{i\omega-i\tilde{\omega}_{n}}-\sum_{n=1}^{N}\frac{\tilde{R}_n^*}{i\omega-i\tilde{\omega}_{n}^*}
\end{equation}
From such a rational fit, one therefore can easily determine the target poles $\tilde{\omega}_n$, their target residues $\tilde{R}_n$, and a direct term $\tilde{A}$.

Now, from the QNMT Eqs.~(\ref{eq:QNMT}) one can write
\begin{equation}
\tilde{S}_{21}(\omega) =\tilde{C}_{21}+\sum_{n=1}^{N}\frac{\tilde{S}_{21}^{(n)}}{i\omega-i\tilde{\omega}_{n}} ,
\end{equation}
where $\tilde{S}_{21}^{(n)}=\tilde{\bar{S}}_{21}^{(n)}\tilde{C}_{11}+\tilde{\bar{S}}_{22}^{(n)}\tilde{C}_{21}$.
Multiplying with its conjugate and expanding the second-order terms, yields an expansion of the form Eq.~(\ref{eq:targetS21}) with
\begin{equation}\label{eq:qnmtS21}
\begin{gathered}
\tilde{A} = |\tilde{C}_{21}|^2\\
\tilde{R}_n = \tilde{S}_{21}^{(n)}\left(\tilde{C}_{21}^*+\sum_{m=1}^{N}\frac{\tilde{S}_{21}^{(m)*}}{i\tilde{\omega}_m^*-i\tilde{\omega}_n}\right).
\end{gathered}
\end{equation}
The first line implies that $\tilde{C}_{21}=\tilde{C}_{12}=\sqrt{\tilde{A}}$ (choosing $\theta=0$), thus $\tilde{C}_{11}=-\tilde{C}_{22}^*=\sqrt{1-\tilde{A}}e^{i\varphi}$ (where again $\varphi$ can be chosen arbitrarily). Moreover, from the QNMT Eq.~(\ref{eq:QNMT}), the terms $\tilde{\bar{S}}_{pq}^{(n)}$ are given as functions of $\{\tilde{\omega}_n,\tilde{\sigma}_n\}$, so the second line is a nonlinear system of $N$ complex equations for $\tilde{\sigma}_n$. However, for a physically realizable system, one also needs to ensure reciprocity via $\tilde{S}_{21}^{(n)}=\tilde{S}_{12}^{(n)}$, therefore, one needs to solve a system of $2N$ complex equations to find the target $\tilde{\sigma}_n$.

\begin{figure}[ht!]
    \centering
    \includegraphics[width=1\linewidth]{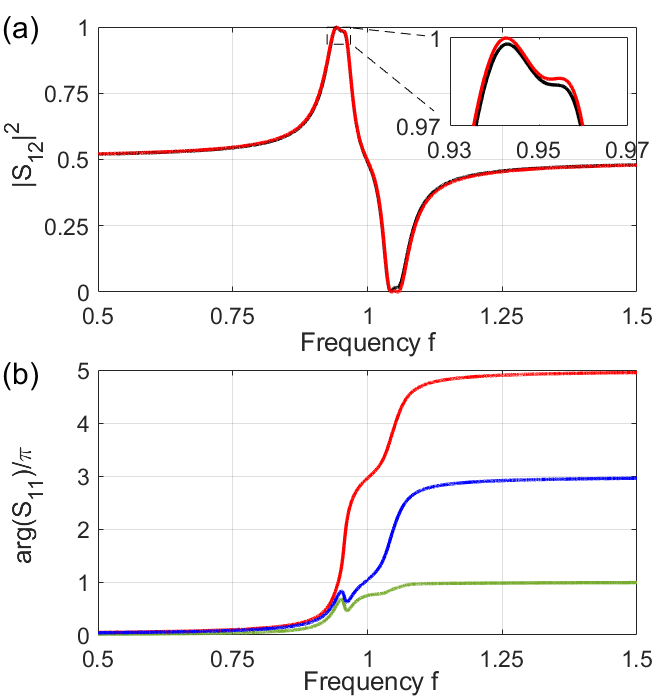}
    \caption{(a) Desired transmission amplitude spectrum (black), generated using $\{f_n\}=\{0.945-0.02i,0.96-0.015i,1.04-0.015i,1.055-0.02i\}$, $\{\sigma_n\}=\{-i,1.5i,i,-1.5i\}$, and ${C}_{21}=\sqrt{0.5}$. The rational fitting procedure exactly recovered the poles and the direct term, while three different solutions were produced for $\{\tilde{\sigma}_n\}=\{-0.08-0.84i,-0.05+0.74i,-0.26+1.07i,0.10-0.85i\}$ (red), $\{0.64-1.11i,-0.60+1.33i,-3.12-3.01i,-4.36+0.09i\}$ (green), $\{0.11-1.17i,0.10+1.34i,0.21+0.88i,-0.14-1.16i\}$ (blue), which all fit well the amplitude spectrum, satisfy reciprocity, but have different reflection-phase responses (b).}
    \label{fig:QNMTfit}
\end{figure}

As an example, consider the arbitrary choice of transmission-amplitude spectrum shown in Fig.~\ref{fig:QNMTfit}(a) with a black line. It was generated via the QNMT Eq.~(\ref{eq:QNMT}), using an arbitrarily chosen set of 4 poles and ratios, and background-transmission amplitude equal to $\sqrt{0.5}$. Interestingly, the chosen generating ratios do \emph{not} satisfy reciprocity exactly (the phases of the generated $\tilde{S}_{21}(\omega)$ and $\tilde{S}_{12}(\omega)$ are slightly different), so we may not be able to fit this spectrum precisely. We followed the procedure above: We sampled this spectrum at 10,000 points and the AAA algorithm (via Julia's BaryRational.jl package~\cite{Bary}) fit it to provide the direct term $\tilde{A}$, poles $\{\tilde{\omega}_n\}$, and residues $\{\tilde{R}_n\}$. For the direct term and poles, the fit recovered exactly the generating values. Then, we solved the nonlinear system of $2N$ complex equations to find, for three different sets of initial guesses, three different solutions for $\tilde{\sigma}_n$ (see Fig.~\ref{fig:QNMTfit} caption), which give the fit residues and satisfy reciprocity. They all provided the same extremely good (though indeed not exact) fit to the desired amplitude spectrum, exhibiting the same transmission-phase response, but differing in the reflection-phase response (which was not specified), as seen in Fig.~\ref{fig:QNMTfit}(b). Any of these sets can now be used as targets for our optimization method to produce a structure with this transmission-amplitude response.

\section{Underdetermined Levenberg-Marquardt algorithm}\label{appendix:LMA}

Recall that to minimize $||\boldsymbol{f}(\boldsymbol{x})||^2$, the Gauss--Newton method at the $i$-th iteration finds the update $\boldsymbol{x}_i+\boldsymbol{\delta}_i$ to the current value $\boldsymbol{x}_i$  by minimizing the residual under the first-order approximation $||\boldsymbol{f}(\boldsymbol{x}_i)+\boldsymbol{J}(\boldsymbol{x}_i)\boldsymbol{\delta}_i||^2$ (where $\boldsymbol{J}(\boldsymbol{x})=\frac{\partial \boldsymbol{f}}{\partial \boldsymbol{x}}$ is the Jacobian), whose solution for $\boldsymbol{\delta}_i$ can be found using the standard normal equation  of the ordinary least square:
\begin{equation}\label{eq:normal_ols}
\boldsymbol{J}(\boldsymbol{x}_i)^T\boldsymbol{J}(\boldsymbol{x}_i) \cdot \boldsymbol{\delta}_i=-\boldsymbol{J}(\boldsymbol{x}_i)^T\boldsymbol{f}(\boldsymbol{x}_i)   
\end{equation}
The LMA~\cite{fletcher2000practical} simply adds a damping/regularization term to Eq.~\ref{eq:normal_ols}:
\begin{equation}\label{eq:normal_LMA}
[\boldsymbol{J}(\boldsymbol{x}_i)^T\boldsymbol{J}(\boldsymbol{x}_i)\boldsymbol+\lambda_i  \boldsymbol \Delta] \cdot \boldsymbol{\delta}_i=-\boldsymbol{J}(\boldsymbol{x}_i)^T\boldsymbol{f}(\boldsymbol{x}_i)   
\end{equation}
where the interpolation parameter $\lambda_i$ changes dynamically on each iteration (and the hyperparameter $\lambda_0$ is chosen as $0.1$ in all our examples) and $\boldsymbol{\Delta}$ is a diagonal matrix (another hyperparameter), which is typically chosen as either simply the identity matrix $\boldsymbol{I}$ or as the problem-scaled matrix $\boldsymbol \Delta=\text{diag}[\boldsymbol{J}(\boldsymbol{x}_i)^T\boldsymbol{J}(\boldsymbol{x}_i)]$.


Usually, the nonlinear least-square problems involve more equations than the number of variables (\#rows of $\boldsymbol{J}\gg$ \#columns of $\boldsymbol{J}$), so the system is overdetermined. However, in the case of TopOpt, there are usually many more variables than equations (\#rows of $\boldsymbol{J}\ll$ \#columns of $\boldsymbol{J}$), making the system underdetermined. Therefore, the matrix $\boldsymbol{J}^T\boldsymbol{J}$ is large and rank-deficient, making the regular LMA formulation of Eq.~(\ref{eq:normal_LMA}) expensive and sometimes (when $\lambda_i\ll1$) even unstable to solve.

Alternatively, the conventional approach for an underdetermined system is to find the minimal-norm solution, which results in the following modified normal equation of Eq.~\ref{eq:normal_ols}:
\begin{equation}\label{eq:modified_ols}
\begin{split}
\boldsymbol{J}(\boldsymbol{x}_i)\boldsymbol{J}(\boldsymbol{x}_i)^T &\cdot \boldsymbol{z}=-\boldsymbol{f}(\boldsymbol{x_i}) \\
\boldsymbol{\delta}_i=&\boldsymbol{J}(\boldsymbol{x}_i)^T\cdot\boldsymbol{z}
\end{split}
\end{equation}
The matrix $\boldsymbol{J}\boldsymbol{J}^T$ is instead small and generically full rank. Using this logic, we made a similar modification to Eq.~\ref{eq:normal_LMA} in LMA to account for the damping/regularization term:
\begin{equation}\label{eq:modified_LMA}
\begin{split}
[\boldsymbol{J}(\boldsymbol{x}_i)\boldsymbol{J}(\boldsymbol{x}_i)^T&+\lambda_i\boldsymbol \Delta] \cdot \boldsymbol{z}=-\boldsymbol{f}(\boldsymbol{x_i}) \\
\boldsymbol{\delta}_i&=\boldsymbol{J}(\boldsymbol{x}_i)^T\cdot\boldsymbol{z}
\end{split}
\end{equation}
In order to make the regularization-term $\boldsymbol{\Delta}$ insensitive to the overall scaling of $\boldsymbol{f}$, we chose $\boldsymbol{\Delta}=\|\boldsymbol{J}(\boldsymbol{x}_i)\|_F^2/\text{length}(\boldsymbol{x})\cdot \boldsymbol I$, where $\|\cdot\|_F$ is the Frobenius norm, which guarantees positive-definiteness for nonzero~$\boldsymbol{J}$.

\section{Transfer-matrix method review and its gradient calculation}\label{appendix:TM}

\subsection{Dielectric 1D-layered stack~\cite{Yeh2005}}

Referring to Fig.~\ref{fig:QNMT}, the backwards transfer matrix $T$ of a two-port is defined as $\begin{pmatrix} s_{+1} \\ s_{-1} \end{pmatrix} = T \begin{pmatrix} s_{-2} \\ s_{+2} \end{pmatrix}$, with power-normalized $|s_{\pm p}|^2=1$. For a 1D-stack of isotropic dielectric layers, for propagation through layer $k$ of refractive index $n_k=\sqrt{\varepsilon_k\mu_k}$ ($\varepsilon_k, \mu_k$ are the relative permittivity and permeability) and thickness $d_k$, and for an interface between layers $k$ and $k+1$, the transfer matrix can be written as
\begin{equation}\label{eq:T_layer}
\begin{split}
    T^{(k)} =& \begin{pmatrix} e^{-ik_o n_k d_k} & 0 \protect\\ 0 & e^{ik_o n_k d_k} \end{pmatrix}\\ 
    T^{(k,k+1)} =& \frac{1}{2\sqrt{\rho_{k,k+1}}}\begin{pmatrix} 1+\rho_{k,k+1} & 1-\rho_{k,k+1} \protect\\ 1-\rho_{k,k+1} & 1+\rho_{k,k+1} \end{pmatrix},
\end{split}
\end{equation}
where $\rho_{k,k+1}=X_{k+1}/X_k$, and $X_k=\sqrt{\varepsilon_k/\mu_k}$ for $E_z$ polarization or $X_k=\sqrt{\mu_k/\varepsilon_k}$ for $H_z$ polarization. The total transfer matrix of a stack of $K$ layers is then
\begin{equation}\label{eq:T_total}
    T = T^{(1)}T^{(12)}T^{(2)}\cdot\cdot\cdot T^{(K-1)}T^{(K-1,K)}T^{(K)}.
\end{equation}


For use in the design criteria Eqs.~(\ref{eq:zeros}) and (\ref{eq:C-constraint}), the transformation to the scattering matrix $S$ is
\begin{equation}\label{eq:TtoS}
    S = \frac{1}{T_{11}}\begin{pmatrix} T_{21} & 1 \\ 1 & -T_{12} \end{pmatrix}.
\end{equation}
Note that reciprocity demands $\text{det}(T)=1 \Leftrightarrow S^T = S$ at all (even complex) frequencies.

Note also that transfer matrices can sometimes encounter numerical issues from ill-conditioning, when $|\text{Im}\{k_o n_k d_k\}|\gg 1$ for some layer $k$ (so $T_{11}^{(k)}\gg1$ or $T_{22}^{(k)}\gg1$) or when the transmission through a set of layers becomes very small (so $T_{11}=1/S_{21}\gg1$). This is why an alternative scattering-matrix formalism is generally more robust~\cite{Li1996Smatrix} and can also be used for optimization~\cite{GhebrebrhanBe09}. Nevertheless, for the problem we examined here, with normal incident waves and thickness-limited layers, the T-matrix method did not encounter problems, and was simpler to implement.

To calculate the gradient with respect to the layer thicknesses $\{d_k\}$, it is easy to calculate $dT/dd_k$, as only the element $T^{(k)}$ in Eq.~(\ref{eq:T_total}) depends on $d_k$. Note that, in this case of transfer matrices, the adjoint method for calculating the derivative is essentially equivalent to saving all intermediate matrices $T^{(1...k-1)}$ during the forward $k=1...K$ matrix multiplications in Eq.~(\ref{eq:T_total}), and then calculating $dT/dd_k$ backwards for $k=K...1$ and evaluating only \emph{once} each matrix $T^{(k+1...K)}$~\cite{Birge2007}.

\subsection{Electronic $LC$-ladder circuit~\cite{jarry2007rf}}

The $\mathcal{ABCD}$ matrix $\mathcal{T}$ of a two-port electronic circuit is defined as $\begin{pmatrix} v_1 \\ i_1 \end{pmatrix} = \begin{pmatrix} \mathcal{A} & \mathcal{B}\\\mathcal{C} & \mathcal{D} \end{pmatrix} \begin{pmatrix} v_2 \\ -i_2 \end{pmatrix}$. Note that, although this $\mathcal{T}$ matrix relates voltage/current at port 1 to those at port 2, thus can also be cascaded, it is \emph{not} the system transfer matrix $T$, which relates power-normalized inputs/outputs and requires knowledge of the input/output impedances. For an $LC$-ladder circuit, consisting of alternating series $L$--$C$ and shunt parallel $L\parallel C$ sections, the respective $\mathcal{ABCD}$ matrices are
\begin{equation}\label{eq:ABCD_LC}
\begin{split}
    \mathcal{T}^{\text{s}(k)} =& \begin{pmatrix} 1 & L_k s+1/C_k s \protect\\ 0 & 1 \end{pmatrix}\\ 
    \mathcal{T}^{\text{p}(k)} =& \begin{pmatrix} 1 & 0 \protect\\ C_k s + 1/L_k s & 1 \end{pmatrix},
\end{split}
\end{equation}
where $s=-i\omega$ (using the physics notation for consistency throughout the article). The total $\mathcal{ABCD}$ matrix of an alternating ladder of $K$ sections, starting with a series section, is then
\begin{equation}\label{eq:ABCD_total}
    \mathcal{T} = \mathcal{T}^{\text{s}(1)}\mathcal{T}^{\text{p}(2)}\mathcal{T}^{\text{s}(3)}\cdot\cdot\cdot \mathcal{T}^{(K-1)}\mathcal{T}^{(K)}.
\end{equation}

If the circuit is connected at its two ports to a generator and a load with impedances $R_\text{g}$ and $R_\text{l}$ respectively, then the scattering matrix $S$ is
\begin{equation}\label{eq:MtoS}
    S = \frac{1}{a+b+c+d}\begin{pmatrix} a+b-c-d & \!\!\!2 \\ 2 & \!\!\!-a+b-c+d \end{pmatrix}
\end{equation}
where $a=\mathcal{A}\sqrt{R_\text{l}/R_\text{g}}$, $b=\mathcal{B}/\sqrt{R_\text{g}R_\text{l}}$, $c=\mathcal{C}\sqrt{R_\text{g}R_\text{l}}$, and $d=\mathcal{D}\sqrt{R_\text{g}/R_\text{l}}$. From reciprocity, $\text{det}(\mathcal{T})=1$ and $S^T = S$. (Note that the circuit is ``symmetric'' if $S_{11}=S_{22}\Leftrightarrow a=d \Leftrightarrow \mathcal{A}R_\text{l}=\mathcal{D}R_\text{g}$.)

The gradient of $\mathcal{T}$ (and then $S$) with respect to all $L_k$, $C_k$ elements is calculated easily, similarly to the previous subsection.

\end{document}